\documentclass[sigconf, 10pt, nonacm]{acmart}
\usepackage{xspace}

\title{A Backward-Compatible Protocol Upgrade for HotNets}
\subtitle{Rethinking HotNets for an Era of Hypergrowth and AI}

\author{Paolo Costa}
\affiliation{%
  \position{HotNets 2026 Co-Chair}
  \institution{Microsoft Research}
  \city{}
  \country{}
}

\author{Michael Schapira}
\affiliation{%
  \position{HotNets 2026 Co-Chair}
  \institution{Hebrew University of Jerusalem}
    \city{}
  \country{}
}

\newcommand{\mypar}[1]{\noindent\textbf{#1.}\xspace}
\newcommand{\tinyskip}{\vspace{0.5em}}

\begin{document}

\maketitle

\section{Introduction}
This document outlines the changes adopted for ACM HotNets 2026, spanning its scope, review process, and program structure. Rather than isolated adjustments, these changes form a coherent effort to clarify and extend HotNets' role as a venue for agenda-setting research, community discussion, and experimentation with how the networking community evaluates, disseminates, and discusses research. In particular, HotNets 2026 broadens its scope to include perspective and community-facing contributions, introduces distinct evaluation criteria for technical and perspective papers, adopts a more collaborative and discussion-oriented review process, rethinks how accepted work is presented and discussed at the workshop, and explores responsible uses of generative AI (GenAI) in reviewing and research dissemination.

We believe these changes will help HotNets continue to serve as a home for ambitious, unconventional, and thought-provoking ideas, while also positioning it as a venue for experimenting with new approaches and formats that larger conferences, e.g., SIGCOMM or NSDI, might later adopt. We use this document to solicit feedback from the community, both on these changes and on how HotNets can best serve the networking community in the future. We plan to collect feedback during and after the event and to prepare a follow-up report summarizing the community's reactions and lessons learned.

% The remainder of this document explains the motivations behind these changes and describes how they are reflected in the design of HotNets 2026.

\section{Protocol specification}

HotNets occupies a distinct role in the SIGCOMM ecosystem. For more than two decades, it has provided a home for unconventional thinking, intellectual risk-taking, and debate about the future of networking.

However, the environment in which HotNets operates has changed substantially. Submission volumes have grown substantially, placing increasing pressure on reviewers and workshop formats designed for smaller communities. At the same time, networking research is evolving rapidly, shaped by new technologies, changing research incentives, growing interest in how the community conducts and evaluates research, and the emergence of generative AI. Taken together, these developments motivate a re-examination of how HotNets can best identify, evaluate, and discuss the ideas that will shape the future of networking. Such reflection is not unique to HotNets: recent calls have urged the broader community to rethink how conferences are organized~\cite{shenker2022rethinking}.

Our objective is to adapt the workshop's scope, processes, and structure to the changing landscape while preserving and strengthening the qualities that have made HotNets a home for ambitious, thought-provoking, and agenda-setting ideas. Consistent with the spirit of HotNets and the scientific method itself, we view this year's workshop as an experiment: an opportunity to put these ideas into practice and learn from the outcome. In particular, for this edition of HotNets, we focus on four goals:

\tinyskip

\mypar{G1: Broadening the scope to include community-facing contributions} 
We view HotNets not only as a venue for ambitious technical ideas, but also as a forum where the networking community can reflect on its own evolution. Many consequential questions, such as ``\textit{Whether/how should GenAI be used for reviewing?}'', ``\textit{Is too much of our work drifting toward short-term industry questions?}'', ``\textit{What should we be teaching the next generation of networking researchers and practitioners?}'', and ``\textit{Are current publication and incentive structures encouraging the kinds of research our community most values?}'', do not fit cleanly within traditional conference venues. As a workshop that combines publication with in-person interaction, HotNets is uniquely positioned to host such discussions. We intend to create space for important community-facing discussions that currently lack a natural home, while preserving HotNets' role as a venue for novel technical contributions.

\tinyskip

\mypar{G2: Accommodating distinct modes of contribution} HotNets' value lies between two undesirable extremes: mature ``mini-SIGCOMM'' papers already on a conventional conference trajectory and provocative but weakly substantiated ideas. Its sweet spot is early, unconventional, and exploratory work that presents a compelling and sound technical or intellectual argument and opens new directions for future research: work that can inspire many subsequent papers rather than merely serving as a compressed version of a conventional conference paper. At the same time, not all influential contributions take the form of new technical results. Some shape a field by synthesizing emerging directions, challenging assumptions, identifying overlooked questions, articulating new research agendas, or reframing how existing systems, practices, and trends are understood. Our goal is therefore both to clarify the level of substance expected of all HotNets papers and to recognize distinct modes of contribution that require different evaluation criteria.

\tinyskip

\mypar{G3: Scaling while maintaining quality}
We must address the workshop's growth, which is increasingly straining both the review process and the event itself. Rising submission volumes place heavy burdens on reviewers, leading to concerns about superficial or inconsistent evaluations. At the same time, larger programs often result in compressed schedules with many short talks and parallel tracks, limiting meaningful presentation and discussion. Without changes, these pressures risk degrading both review quality and the workshop experience. A key goal is therefore to manage this scale while preserving the depth and quality that define HotNets.

\tinyskip
\mypar{G4: Engaging with the opportunities of generative AI}
Other communities, particularly in machine learning and AI, are already experimenting with the use of large language models (LLMs) in authoring, reviewing, and artifact evaluation. A randomized study at ICLR 2025, for example, found that LLM-generated feedback can improve the quality of paper reviews~\cite{thakkar2025llmreview}. The networking community, by contrast, has so far made fewer such experiments at the venue level. We aim to proactively explore how generative AI can be incorporated into the evaluation, dissemination, and discussion of research in ways that strengthen, rather than undermine, the community's standards and practices.

\section{Protocol implementation}

The following changes to HotNets are designed to address each of the four goals outlined above in a coherent and mutually reinforcing way.

\subsection{Widening the pipe: extending the call for papers}

We explicitly broaden the scope of HotNets to include not only novel technical ideas, but also substantive community-facing and meta-level contributions, e.g., on how the community conducts and evaluates research, trains the next generation, and structures its publication and incentive systems in response to technological and societal change.

HotNets is a natural venue for such contributions in the form of perspectives, proposals, critiques, and empirical lessons, complementing its longstanding role as a home for early-stage technical ideas.

Broadening the scope does not mean lowering standards. We expect all HotNets papers, whether technical or community-facing, to be original, intellectually provocative, well-grounded, and capable of stimulating meaningful discussion within the community. A procedural or purely descriptive treatment of community topics is not sufficient.

\subsection{DiffServ: recognizing two distinct modes of contribution and evaluation}

We recognize that impactful contributions can take different forms and should be evaluated accordingly. Some papers advance the field through new technical or empirical results, while others do so by synthesizing emerging directions, challenging assumptions, articulating new research agendas, or offering new ways of understanding important technical or community questions.

To reflect this distinction, we introduce two submission categories:

\begin{enumerate}

\item \textbf{Technical papers:} papers whose primary contribution is a new technical idea, system, analysis, methodology, empirical finding, or other concrete research result. We explicitly encourage bold, early-stage ideas that may not yet be fully validated but are technically sound, well-reasoned, and sufficiently grounded to support meaningful engagement by the community. Full implementations or exhaustive evaluations are not required.

\vspace{0.05in}
\item \textbf{Perspective papers:} papers whose primary contribution is a new way of understanding, critiquing, synthesizing, or prioritizing important technical or community-facing questions. These may address technical research directions, methodologies, evaluation practices, education, publication and reviewing processes, or broader questions about the evolution of the networking community. They will be evaluated based on the originality and substance of the perspective, the strength of the argument, and their potential to influence future research and community practices.
\end{enumerate}

Topic and contribution type are separate axes. A paper's topic may be technical or community-facing, and, independently, its primary contribution may be a new technical result or a new perspective. A perspective paper may address a technical research direction, and a technical paper may target a community-facing issue (e.g., a new evaluation methodology or a system that enables data-driven research on previously inaccessible questions).

\subsection{Managing congestion: redesigning review and program structure for scale}

HotNets exists to identify promising early-stage ideas, not to reward consensus around mature work. Together, the changes below redesign the review process to focus attention on a paper's core intellectual contribution, encourage meaningful reviewer discussion, and better align reviewing with HotNets' mission.

\tinyskip

\mypar{Collaborative reviewing}
Building on the approach adopted at NINeS 2026~\cite{nines2026}, we introduce a review process centered on structured discussion and synthesis. Reviewers will first provide concise independent assessments, intended primarily as input to the reviewer discussion (``reviews for reviewers''), followed by discussion and a designated reviewer's synthesis capturing the main points of agreement and disagreement. This synthesis will serve as the primary feedback to authors, providing a clear and coherent summary while preserving diverse perspectives. The goal is to enable a more frank and open conversation among reviewers and reduce the burden on individual reviewers, while providing authors with more coherent and structured feedback.

\tinyskip

\mypar{Adopting a binary, sufficient support model}
To improve scalability and create room for more controversial and thought-provoking ideas, we adopt a ``sufficient support'' model: a paper is accepted if, after discussion, it receives explicit support from at least two reviewers. This approach, explored previously at HotNets 2023~\cite{hotnets2023}, reflects the view that if a meaningful subset of the reviewers sees clear value in a submission, it should be accepted and put before the community. While this approach may increase the number of \emph{false positives} (papers incorrectly accepted), we believe the risk of \emph{false negatives} (papers incorrectly rejected) is more detrimental to the community. Consistent with this philosophy, HotNets 2026 will move away from fine-grained numerical scores toward a simpler binary model centered on support for acceptance. This aligns with HotNets' goal of surfacing early, potentially controversial ideas while reducing the tendency toward middle-of-the-scale consensus in conventional reviewing. A natural consequence is that this model tends to enlarge, rather than shrink, the accepted set. We address this not by capping acceptances, but through the discussion-oriented program described below, which scales to more papers than a schedule built around sequential talks while preserving depth of engagement.

\tinyskip

\mypar{Evaluating core intellectual contributions first}
The review process will begin with an assessment of the paper's central contributions, intellectual ambition, and potential to influence future research or stimulate discussion within the HotNets community. Consistent with HotNets' focus on early, agenda-setting ideas, particular emphasis will be placed on the abstract and introduction, which should clearly articulate the paper's key contributions and why they matter. This initial assessment will determine whether the paper should be further discussed and evaluated. If \emph{all} reviewers agree that a paper does not have sufficient core contributions, it may be rejected without further review. This approach allows reviewers to focus their attention on papers with the most potential impact, while also providing authors with early feedback on the clarity and significance of their work.

\tinyskip

\mypar{Asynchronous reviewing}
The above changes naturally support an asynchronous end-to-end review process, allowing many decisions to emerge from reviewer support and discussion without lengthy, often inefficient, meetings. Accordingly, we will not convene a full online Technical Program Committee (TPC) meeting. Instead, deliberation will take place primarily through per-paper discussions on HotCRP, with targeted synchronous conversations used only when additional discussion is needed. This approach promotes deeper engagement from the most relevant reviewers while keeping overall overhead low.

\tinyskip

\mypar{A discussion-oriented program structure}

All accepted papers will appear in the HotNets proceedings and will be integrated as central components of the workshop program.

We believe the traditional HotNets workshop model does not scale well to the kind of deep technical engagement and discussion that HotNets is intended to foster. Many uniform short talks often lead to fragmented, difficult-to-follow sessions with limited opportunity for sustained interaction, while parallel tracks risk weakening the shared workshop experience. Instead, we will structure the workshop around interaction and discussion. Interactive, paper-focused sessions, using posters, demos, breakout discussions, and related formats, will play a central role in the program. These formats create space for deeper technical engagement, detailed feedback, and sustained discussion around individual papers.

In parallel, the workshop will feature curated plenary sessions organized around broader themes, tensions, or emerging directions reflected across submissions. These may include cross-paper discussions, panels, debates, thematic talks, or related interactive formats involving both accepted papers and invited participants. Participation in these sessions will be determined by the potential of the topic or collection of papers to stimulate broad and productive discussion.

The goal is to shift the emphasis of the workshop from presentation toward engagement: creating a program structure that better supports interaction, debate, synthesis across ideas, and collective reflection on important technical and community questions. This approach is reminiscent of the flipped classroom model~\cite{lage2000inverting, bishop2013flipped}, where the workshop serves as a forum for discussion and exploration of ideas rather than a venue for one-way presentation.

\subsection{The eighth layer: engaging with GenAI in reviewing and dissemination}

\tinyskip

\mypar{AI-assisted reviewing}
We believe the community needs a deeper and more systematic discussion about how generative AI can, and should, be integrated into the reviewing process. While there are clear risks around reliability, bias, confidentiality, and overreliance, we are also optimistic that LLMs can become a valuable tool for improving review quality~\cite{thakkar2025llmreview} and reducing reviewer burden when used carefully and with appropriate safeguards.

As a first step, HotNets 2026 adopts a conservative and reviewer-centered approach in which LLMs assist reviewers rather than evaluate papers themselves. AI tools may be used to help reviewers identify related literature, surface potential factual inconsistencies or omissions in reviews (e.g., when a review claims a paper does not discuss something that is in fact addressed), suggest alternative interpretations or questions for consideration, and improve the clarity or completeness of review text. To preserve confidentiality, any such use must rely only on LLM services that guarantee zero data retention (i.e., submissions and reviews are neither stored nor used for training), and every AI-generated suggestion must be verified against the paper before it is acted upon. This assistance is also opt-in: it applies only to submissions whose authors have explicitly agreed to it. Under this model, LLMs do not score, rank, or judge papers, and they do not replace reviewer expertise or discussion. Responsibility and accountability for reviews and decisions remain fully with human reviewers and program chairs.\footnote{In the same spirit, the authors consulted an LLM for feedback while preparing this document. Consistent with the principle above, all responsibility and accountability for its content remain with the (human) authors.} The role of AI is solely to support reviewers in producing more careful, informed, and comprehensive evaluations.

\tinyskip

\mypar{Encouraging rich, AI-enabled supplementary artifacts}
We encourage authors of accepted papers to complement their work with additional materials, such as recorded talks, slides, extended reports, and demos or other interactive artifacts. This reflects a broader shift in how research is consumed: a six-page paper is often only one entry point into a contribution, and different formats can make ideas more accessible to a wider audience.

We particularly highlight the opportunity to use emerging AI-based tools (e.g., LLM-based notebooks or LLM-generated artifacts) to enable richer interaction with accepted work. These tools can support alternative forms of engagement, such as generated summaries at different levels of detail, audio or podcast-style explanations, video summaries, and interactive exploration of a paper's content. Such artifacts can help a broad audience quickly grasp unfamiliar topics, revisit ideas over time, and engage more deeply with the work. Our goal is not to mandate new formats, but to encourage experimentation with complementary ways of communicating research.

\section{Conclusion}

HotNets has long been a place where the networking community explores ambitious and unconventional ideas. We believe this spirit should extend not only to the technical directions pursued by the community, but also to how research is evaluated, discussed, and disseminated. The changes described above are intended to preserve HotNets' distinctive culture of intellectual risk-taking while adapting it to a growing community and a rapidly changing research landscape. Our goal is for HotNets to remain not only a venue for exploring the future of networking, but also a place where the networking community can explore its own future, both by discussing it and by experimenting with it. We look forward to engaging with the community on these changes and to learning from the experience of HotNets 2026. 

\begin{acks}
We thank Gianni Antichi, Katerina Argyraki, Behnaz Arzani, Ignacio Castro, Jon Crowcroft, Nate Foster, Brighten Godfrey, Nick McKeown, Aurojit Panda, Peter Pietzuch, 
Barath Raghavan, Costin Raiciu, Scott Shenker, and Keith Winstein for their valuable feedback on this manuscript.
\end{acks}

\bibliographystyle{ACM-Reference-Format}
\bibliography{rethinking_hotnets_2026}

\end{document}